\documentclass[12pt]{iopart}
\usepackage{amsmath}
\usepackage{iopams}  
\usepackage{graphicx}

\begin{document}

\title{Quadrupolar power radiation by a binary system in a hyperbolic encounter on de Sitter background }

\author{Michael Blanc}
\address{ETH Z\"urich, Switzerland  and } 
\address{ Department of Physics, University of Z\"urich, Winterthurerstrasse 190, 8057 Zurich, Switzerland}
\author{Philippe Jetzer} 
\address{ Department of Physics, University of Z\"urich, Winterthurerstrasse 190, 8057 Zurich, Switzerland}
\author{Shubhanshu Tiwari}
\address{ Department of Physics, University of Z\"urich, Winterthurerstrasse 190, 8057 Zurich, Switzerland}

\begin{abstract}

    The present cosmological model and the surveys favor the universe with a small but positive cosmological constant $\Lambda$,  which accounts for dark energy and causes an exponential expansion. This can have observational consequences in the current detection of gravitational waves, as most of the waveforms for gravitational radiation are computed assuming a flat (Minkowski) background. In this work, we compute gravitational radiation within the quadrupole approximation on positive $\Lambda$ (de Sitter) background for a binary system interacting gravitationally through a hyperbolic encounter. 
    We quantify the influence of the cosmological constant on the radiated energy as small corrections to the leading order Minkowski background results. The first order de Sitter background correction is of the order $\sqrt{\Lambda}$, and is thus extremely small. 
    Therefore, the cosmological constant influence on the gravitational radiation is negligible and may not be detected with the existing or planned gravitational wave detectors. 
\end{abstract}

\section{Introduction}

Routine detection of gravitational wave events by the current generation of ground-based detectors LIGO,Virgo, and KAGRA has opened the field of gravitational wave astronomy \cite{gwtc-3,ligo,virgo,kagra}. With the upcoming space-based detector LISA and the next generation of ground-based detectors, subtle and rare effects will affect the gravitational waveform. Currently, the gravitational waveforms employed in the data analysis of current generation of consider asymptotically flat (Minkowski) spacetime. Cosmological observations since the last few decades have hinted towards a small but positive and nonzero cosmological constant \cite{Sahni:1999gb}, and hence the spacetime will be asymptotically de Sitter. The difference between asymptotically de Sitter and Minkowski spacetime on the gravitational waveforms should be computed to estimate if the bias is due to ignoring asymptotically de sitter spacetime. In this work we present this computation for the hyperbolic encounters of compact objects. 


Hyperbolic encounters are single scattering events where the majority of the energy is released near the point of closest approach \cite{zp-hype,gb-hype}, under the form of gravitational waves. Every physical quantity characterising the hyperbolic encounter can be expressed in terms of only four variables, that are the impact parameter, the initial relative velocity and the masses of the bodies, if we neglect their spins. Several works have considered hyperbolic encounters: reference \cite{cap-hype} determines the energy emitted by a hyperbolic encounter in a flat Minkowski universe, while references \cite{bong-hype} and \cite{ho-hype} investigate the mean power released by a binary system in a de Sitter universe, respectively, on circular and elliptic orbits. In this work, we extend those results to the case of a hyperbolic encounter in a de Sitter universe.
 
The paper is organized as follows. First we briefly describe how to compute the quadrupolar radiation
from hyperbolic trajectories in Minkowski space; and then we discuss the gravitational radiation in a de Sitter space.
Lastly, we present the results for quadrupolar radiation from hyperbolic encounters in a de Sitter background.
We use units in which $c=1$ and the metric signature $(-,+,+,+)$. 
Regarding the index notation, greek letters denote
spacetime indices and range from 0 to 3, whereas latin letters denote space indices and range from 1 to 3.

\section{Quadrupolar radiation by a hyperbolic encounter on a Minkowskian background} \label{hyperbolic_encounter_Minkowski_background}

The quadrupole formula resorts to the time averaging of a time-dependent quantity. In the case of a bounded system, such as a binary system on circular or elliptic orbits, it makes sense to average the power over a complete orbit since it yields the mean power of the continuous gravitational waves. In the case of unbounded orbits, such as a binary system with parabolic or hyperbolic trajectories, the approach is trickier. Indeed, it would not make sense to average the power because the wave emission is not periodic at all. Waves begin to be emitted as soon as both stars are sufficiently close to each other, and the wave intensity is maximal when both stars reach their periastron (closest approach between the two interacting stars). Without the time average, the quadrupole formula gives the instantaneous radiated power $P(t)$ as a function of time. Since the interacting time is finite, a relevant way of quantifying the gravitational waves emitted by a hyperbolic encounter is to calculate the whole radiated energy generated by this encounter through power integration over the entire trajectory.

\begin{figure}[h!]
    \centering
    \includegraphics[width=4cm]{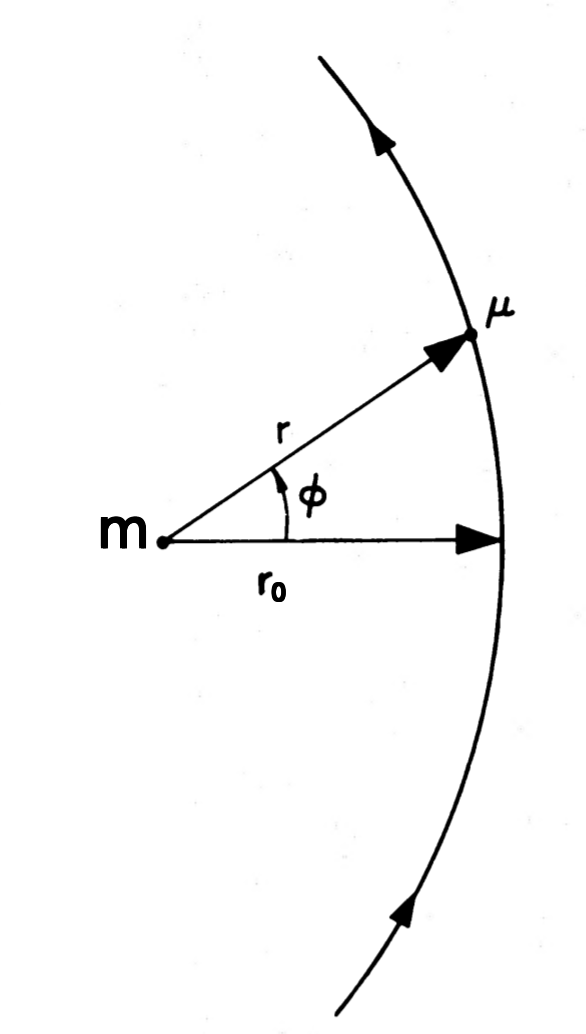}
    \caption{The geometry of a hyperbolic encounter and its most important parameters (see e.g. \cite{Turner_1977}). The reduced mass $\mu$ is moving on a hyperbolic orbit with focus at the origin of the reference frame where the total mass $m$ lies. The motion of $\mu$ is described by the vector radius $r$ and the polar angle $\phi$. The vector radius $r_0$ (corresponding to the polar angle $\phi=0$) represents the periastron distance. $\mu$ is coming from $-\phi_{\infty}$ and ends up at $+\phi_\infty$.}
    \label{hyp_encouter_image_2}
\end{figure}

In this section we briefly review the
spontaneous radiated power and the total released energy in terms of the hyperbola eccentricity $\epsilon>1$ and the periastron distance $r_0$ (another way to do it is to use instead the periastron angle $\phi_0$, the initial relative velocity $v_0$ and the impact parameter $b$). For that purpose, we consider the system depicted in Figure \ref{hyp_encouter_image_2}.
The differential equation for the inverse radius $u=1/r$ is given by
\begin{equation}
    \frac{d^2u}{d\phi^2}+u=\frac{Gm}{L^2}~,
\end{equation}
($L$ being the angular momentum per unit mass, m=($m_1+m_2$) and the reduced mass is $\mu = (m_1 m_2)/m$) whose general solution is of the form
\begin{equation}
    u=A\cos\phi+B\sin\phi+\frac{Gm}{L^2}~.
\end{equation}
The constant $B$ is fixed by imposing a condition on the radial velocity at periastron (dot denotes time derivative)
\begin{equation}
    \dot{r}(\phi=0)=0\quad\Leftrightarrow\quad0=\dot{u}(\phi=0)=\big[\dot{\phi}(B\cos\phi-A\sin\phi)\big]_{\phi=0}=B\,\dot{\phi}(\phi=0)~.
\end{equation}
As $\dot{\phi}(\phi=0)>0$, we infer $B=0$. The constant $A$ is set by the radial condition at periastron
\begin{equation}
    \frac{1}{r_0}=u(\phi=0)=A+\frac{Gm}{L^2}\quad\Leftrightarrow\quad A=\frac{1}{r_0}-\frac{Gm}{L^2},
\end{equation}
hence the radial distance reads
\begin{equation}
    r=\frac{1}{(\frac{1}{r_0}-\frac{Gm}{L^2})\cos\phi+\frac{Gm}{L^2}}=\frac{\frac{L^2}{Gm}}{1+(\frac{L^2}{Gmr_0}-1)\cos\phi}~,
\end{equation}
and gets the following form by defining the eccentricity as
\begin{equation} \label{epsilon_def}
    \epsilon:=\frac{L^2}{Gmr_0}-1\,;  
    ~~~~{r=\frac{r_0(\epsilon+1)}{1+\epsilon\cos\phi}}~.
\end{equation}
The angular momentum $L$ follows from the definition of the eccentricity, $L=\sqrt{Gmr_0(\epsilon+1)}$, and the time derivative of the angle $\phi$ can then be expressed as a function of $\phi$ only
\begin{equation} \label{phi_point}
    \dot{\phi}=\frac{L}{r^2}=\frac{\sqrt{Gm}}{r_0^\frac{3}{2}(1+\epsilon)^\frac{3}{2}}(1+\epsilon\cos\phi)^2~.
\end{equation}
Finally, both asymptotic angles $\pm\phi_\infty$ are determined by imposing the radial condition at infinity:
\begin{equation}
    u(\phi=\pm\phi_\infty)=0\quad\Leftrightarrow\quad1+\epsilon\cos(\pm\phi_\infty)=0\quad\Leftrightarrow
    {\phi_\infty=\arccos\Big(\frac{-1}{\epsilon}\Big)}~.
\end{equation}

We can now compute the radiated power in Minkowski space, where the quadrupole moment tensor 
is given by Einstein's formula. The time derivatives of moments are displayed in appendix \ref{DOM_M2}. 
For the emitted power $P$ we get:
\begin{equation} \label{power_method2_Minkowski}
    P=\frac{4\mu^2m^3G^4\,(1+\epsilon\cos\phi)^4}{15\,r_0^5\,(1+\epsilon)^5}\big[24+13\,\epsilon^2+48\,\epsilon\cos\phi+11\epsilon^2\cos(2\phi)\big]~.
\end{equation}
The total energy emitted in the form of gravitational waves during the hyperbolic encounter is found 
by integrating the emitted power over the entire trajectory
\begin{equation}
    \Delta E=\int_{-\infty}^\infty P\,dt=\int_{-\phi_\infty}^{\phi_\infty}\frac{P}{\dot{\phi}}\,d\phi~,
\end{equation}
where the relation (\ref{phi_point}) has to be taken to get an integrand depending only on $\phi$. The total emitted energy expressed in terms of the eccentricity, the periastron distance and both total and reduced masses is thus given by
\begin{equation}\label{total_energy_Mink_method2}
\begin{split}    
        \Delta E=\frac{2G^\frac{7}{2}\mu^2m^\frac{5}{2}}{45r_0^\frac{7}{2}(\epsilon+1)^\frac{7}{2}}\bigg(3\arccos\Big(\frac{-1}{\epsilon}\Big)(96+292\epsilon^2+37\epsilon^4) \\ +\sqrt{\epsilon^2-1}\,(602+673\epsilon^2)\bigg)~.
\end{split}
\end{equation}
We can easily find the parabolic limit ($\epsilon\rightarrow1$), for which we get:
\begin{equation}  \label{parabolic}
    \Delta E=\frac{85\pi\,G^\frac{7}{2}\mu^2m^\frac{5}{2}}{12\sqrt{2}\,r_0^\frac{7}{2}}~.
\end{equation}

\section{Gravitational radiation in a de Sitter universe} \label{meilleure_section}

\subsection{De Sitter universe} \label{dsitterU}

The de Sitter universe describes a flat universe ($k=0$)  containing only dark energy, as given in terms of a positive cosmological constant $\Lambda>0$. In this particular model, the only non-vanishing energy density is the vacuum energy (or dark energy) density $\rho_\Lambda:=\frac{\Lambda}{8\pi G}$. In the de Sitter universe, the Friedmann equations become:
\begin{equation}
    \frac{\dot{a}^2}{a^2}=\frac{\ddot{a}}{a}=\frac{\Lambda}{3},\quad\dot{\rho}=0~,
\end{equation}
since the cosmological constant is time independent. Both equations involving the scale factor give the same solution:
\begin{equation} \label{scale_factor}
    a(t)=e^{Ht},\quad H:=\frac{\dot{a}}{a}=\sqrt{\frac{\Lambda}{3}}~,
\end{equation}
where $H$ is the Hubble rate and quantifies the expansion of the de Sitter universe, whereas the cosmic time $t\in[-\infty,\infty]$ is chosen such that the Bing-Bang corresponds to $t=-\infty$, and today to $t_0=0$ such that $a(t_0)=1$. The de Sitter universe being flat, its metric therefore simply reads:
\begin{equation} \label{de_Sitter_metric}
    ds^2=-dt^2+e^{2Ht}\big(dr^2+r^2d\theta^2+r^2\sin\theta\,d\phi^2\big)=-dt^2+a(t)^2\,d\mathbf{x}^2~.
\end{equation}
The de Sitter model therefore represents an empty universe expanding forever due to vacuum energy.

\subsection{Energy flux and de Sitter quadrupole formula}
Here we use the results given in Section 4 of \cite{Hoque:2018byx} and Section III.A.1 of \cite{Date:2015kma}.
The expression for the power radiated by an isolated source in a de Sitter background is given by the de Sitter quadrupole formula:
\begin{equation} \label{de_Sitter_quadrupole_formula}
    {P=\frac{G}{5}<R^{ij}R_{ij}-\frac{1}{3}(R^l_l)^2>(t_{ret})}~,
\end{equation}
where the brackets denote time average, $R_{ij}=\partial_t^3Q_{ij}-3H\partial_t^2Q_{ij}+2H^2\partial_tQ_{ij}+H\partial_t^2\Bar{Q}_{ij}-H^2\partial_t\Bar{Q}_{ij}$ is the radiation field, $R^l_l:=\delta^{kl}R_{kl}$, and we recall the energy and pressure quadrupole moments:
\begin{equation}
\begin{split} 
        Q_{ij}(t_{ret})&:=\int a^3(t_{ret})T_{00}(t_{ret},\mathbf{x})x_ix_j\,d^3\mathbf{x}  \\
        \Bar{Q}_{ij}(t_{ret})&:=\int a^3(t_{ret})\eta^{kl}T_{kl}(t_{ret},\mathbf{x})x_ix_j\,d^3\mathbf{x}~,
     \end{split}
\end{equation}
($\eta$ being the Minkowski metric).
For $H=0$, corresponding to the flat and static universe, we get $R_{ij}=\dddot{Q}_{ij}$ and we recover the quadrupole formula of Einstein with $Q_{ij}\rightarrow Q_{ij}-\frac{1}{3}Q^l_l\,\delta_{ij}$. $T_{\mu\nu}$ is the energy-momentum 
tensor.

As expected, the de Sitter quadrupole formula comprises an expansion in powers of $H$, whose 0-th order term is the Minkowskian one. An interesting property about waves generated on de Sitter spacetime is that the power carries information about the energy density and the pressure density, contrary to the Minkowskian case where only the energy density comes into play (to lowest post-Newtonian order) \cite{bong-hype}. Let us recall the assumptions made to arrive at this final de Sitter quadrupole formula:
i) the physical size of the source is much smaller than the cosmological horizon; 
ii) the bodies involved in the source have a velocity which is small compared to the speed of light; 
iii) and the source is only dynamically active for a finite time period.

Note that the TT-tensor is the correct notion of transverse traceless tensors. However, as discussed in \cite{ho-hype} in the context of the power radiated by a circular binary system on a de Sitter background, the result using the tt projection exactly matches with that of the calculation done in paper \cite{bong-hype} using TT extraction. This indicates that for the energy flux computation by a circular binary system, TT versus tt does not matter in de Sitter spacetime. Whether this generalizes to elliptic and/or hyperbolic orbits still remains to be investigated.

\section{Quadrupolar radiation by a hyperbolic encounter on a de Sitter background} 

\paragraph{Assumptions}
With the de Sitter quadrupole formula (\ref{de_Sitter_quadrupole_formula}) we can now compute the radiated energy during the whole hyperbolic encounter. For this we will make the following assumptions:
\begin{itemize}
    \item The characteristic proper time scale of the encounter $t_{en}$ and the expansion rate of the background are assumed to be such that the expansion of the universe can be neglected during the encounter: $a\approx\text{const}=1$, $Ht_{en}\ll1$. Thus, a static de Sitter universe is assumed during the encounter. Freedom in the normalisation of the scale factor $a$ allows us to set it to unity. \\
    \item The pressure of each body is negligible beside their energy density, thus the radiation field reduces to $R_{ij}\approx\partial_t^3Q_{ij}-3H\partial_t^2Q_{ij}+2H^2\partial_tQ_{ij}$. \\
    \item The relative physical separation $R:=ar=a||\mathbf{x}_1-\mathbf{x}_2||$ is such that the bodies are far apart compared to the Schwarzschild radius of either body at any time: $\frac{2Gm}{c^2}\ll R$ and  $\frac{2G\mu}{c^2}\ll R$. \\
    \item Each body moves slowly compared to the speed of light: $v/c \ll1$. \\
    \item The trajectory of the binary is well approximated by a hyperbolic orbit: orbit shrinking steming from energy loss due to gravitational radiation and orbit expansion due to the universe expansion are both neglected.
\end{itemize}
These assumptions can be interpreted as the Newtonian approximation which we use to describe the motion of the binary system. In the following, we denote by a capital $R$ the physical distance while keeping a small $r$ for the comoving distance, and for components $\Bar{x}^i=ax^i$ denotes the physical distance.

\paragraph{Quadrupole moment}
According to references \cite{ho-hype},\cite{bong-hype} the time component of the source energy-momentum tensor and the mass density are linked by:
\begin{equation}
    \begin{split}
        T_{00}(\mathbf{x},t)&=a^2(t)\rho(\mathbf{x},t) \\
        \rho(\mathbf{x},t)&=\mu\,\delta(\Bar{\mathbf{x}}-\Bar{\mathbf{x}}_*(t))=\frac{\mu}{a^3}\,\delta(\mathbf{x}-\mathbf{x}_*(t))~,
    \end{split}
\end{equation}
where $\Bar{\mathbf{x}}_*(t)=a(t)\,\mathbf{x}_*(t)$ denotes the physical trajectory of the body of mass $\mu$. The energy quadrupole moment of the binary system therefore reads:
\begin{equation}
    Q^{ij}=\int a^3T_{00}\,x^ix^j\,d^3\mathbf{x}=\int\mu a^2\delta(\mathbf{x}-\mathbf{x}_*)\,x^ix^j\,d^3\mathbf{x}=\mu a^2\,x^i_*\,x^j_*=\mu\,\Bar{x}^i_*\,\Bar{x}^j_*~,
\end{equation}
so that
\begin{equation} \label{allezlol}
    Q^{ij}=
    \begin{pmatrix}
        \mu R^2\cos^2\phi & \mu R^2\cos\phi\sin\phi \\
        \mu R^2\cos\phi\sin\phi & \mu R^2\sin^2\phi 
    \end{pmatrix}
    =\mu\,r^2
    \begin{pmatrix}
        \cos^2\phi & \cos\phi\sin\phi \\
        \cos\phi\sin\phi & \sin^2\phi 
    \end{pmatrix}~,
\end{equation}
where we used the first among the above assumptions for the last equality.

\begin{figure}[h!]
    \centering
    \includegraphics[width=7cm]{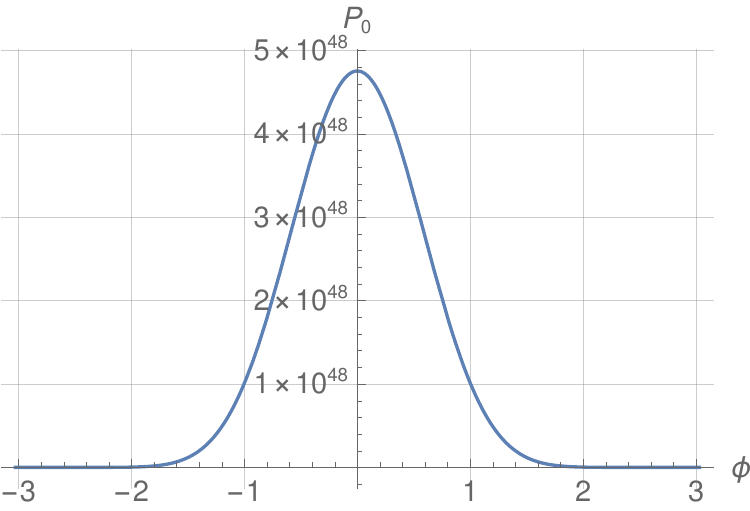}
    \includegraphics[width=7cm]{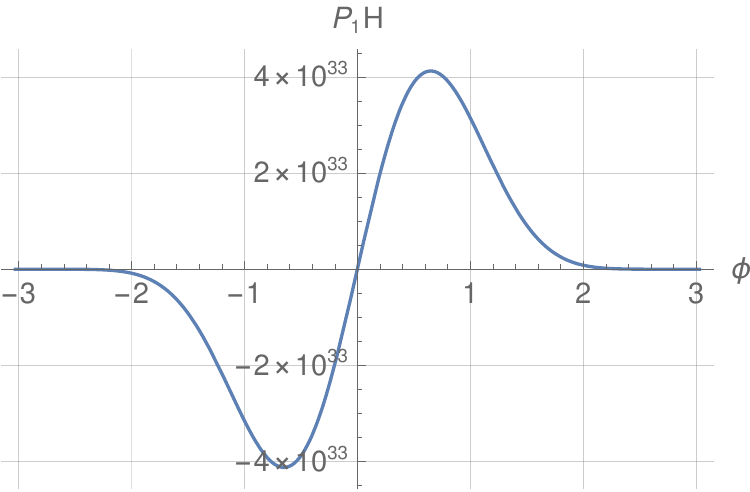}
    \includegraphics[width=7cm]{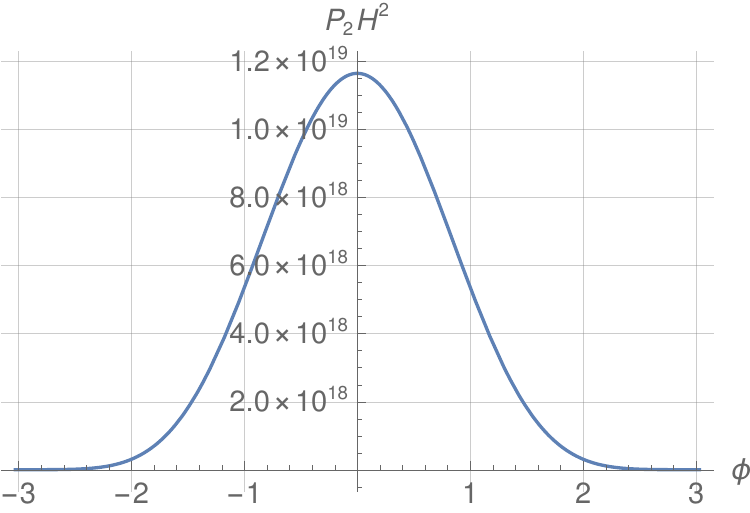}
    \includegraphics[width=7cm]{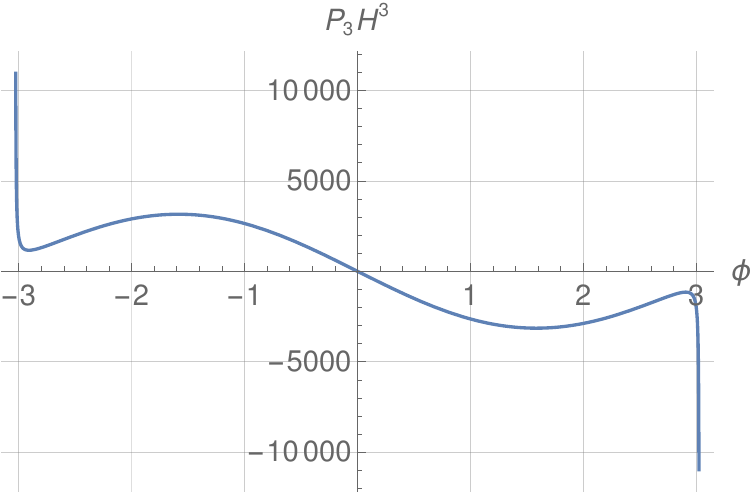}
    \includegraphics[width=7cm]{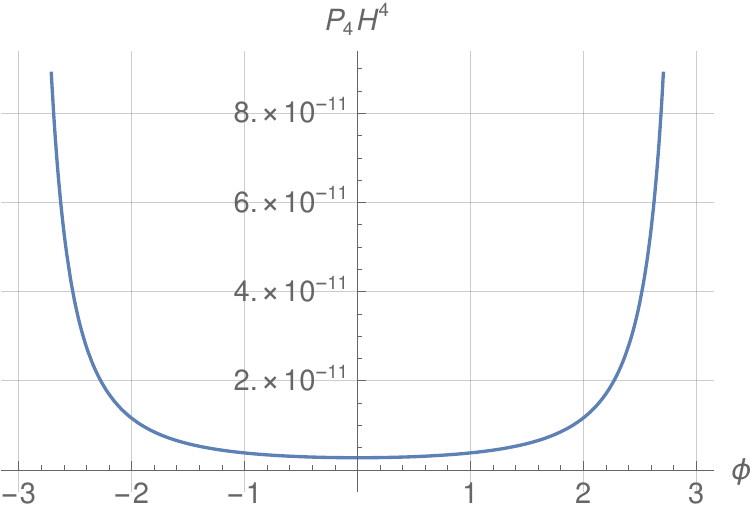}
    \caption{Angular distribution of the power radiated by a hyperbolic encounter of two supermassive black holes with masses $m_1=m_2=0.5 \times 10^{7} {M_\odot}$, impact parameter $b=10 AU$ and initial relative velocity $v_0=10^7 m/s$. The different contributions are displayed. The asymptotic angles are $\phi_\pm=\pm3.02906$. The asymptotes of $P_3H^3$ and $P_4H^4$ respectively reach $\pm1.45\cdot10^{7}$ and $0.48$.}
    \label{power_plot_NA1}
\end{figure}

\paragraph{Power radiation}
In section \ref{hyperbolic_encounter_Minkowski_background}, we derived the trajectory of the reduced mass and the time derivative of the polar angle $\phi$, with $r$ as given by eq. (\ref{epsilon_def}) and $\dot\phi$ by eq. (\ref{phi_point}). To get the radiated power we need the time derivatives of the quadrupole moments, which are given in Appendix \ref{DOM_M2}. We write the total radiated power as:
\begin{equation}
    P = \sum_{n=0}^{4} P_n H^n,
\end{equation}
where the $P_n$ are given by
\begin{equation}
    \begin{split}
        P_0&=\frac{4\mu^2m^3G^4\,(1+\epsilon\cos\phi)^4}{15\,r_0^5\,(1+\epsilon)^5}\big[24+13\,\epsilon^2+48\,\epsilon\cos\phi+11\epsilon^2\cos(2\phi)\big]~, \\
        P_1&=\frac{4G^\frac{7}{2}\mu^2m^\frac{5}{2}}{5(1+\epsilon)^\frac{7}{2}r_0^\frac{7}{2}}(1+\epsilon\cos\phi)^2\sin\phi\Big[18\epsilon+13\epsilon^3+40\epsilon^2\cos\phi+9\epsilon^3\cos(2\phi)\Big]~, \\
        P_2&=\frac{2G^3m^2\mu^2}{15r_0^2(1+\epsilon)^2}\bigg(60+100\epsilon^2+36\epsilon^4+(180+113\epsilon^2)\epsilon\cos\phi+116\epsilon^2\cos(2\phi)+19\epsilon^3\cos(3\phi)\bigg)~, \\
        P_3&=-\frac{32G^\frac{5}{2}\mu^2m^\frac{3}{2}\epsilon^2}{5\sqrt{r_0(1+\epsilon)}}\cdot\frac{\epsilon+\cos\phi}{1+\epsilon\cos\phi}\sin\phi~, \\
        P_4&=\frac{4G^2r_0m\mu^2(1+\epsilon)}{15(1+\epsilon\cos\phi)^2}\Big[6+7\epsilon^2+12\epsilon\cos\phi-\epsilon^2\cos(2\phi)\Big]~.
    \end{split}
\end{equation}
It is straightforward to verify that in the Minkowski case (corresponding to $H=0$ and $P=P_0$) we recover the result given in eq.(\ref{power_method2_Minkowski}). In Figure \ref{power_plot_NA1} we 
plot the angular distribution of the various $P_n$ for a given mass of the colliding black holes as an example. For $P_0$ the emission of gravitational waves is maximal at periastron ($\phi=0$), when the two bodies reach their closest approach, and
the values of $P_0$ are symmetric with respect to $\phi=0$, decreasing as the two bodies move away from each other. 
The odd de Sitter power's contributions ($P_1$ and $P_3$) when integrated over the whole trajectory 
to get the radiated energy do not contribute to it since they are odd with respect to the periastron and thus the
corresponding integral vanishes.
The plot of the leading de Sitter contribution $P_2H^2$ has the same shape as $P_0$, but 29 orders of magnitude smaller. However, the
even much smaller contribution $P_4H^4$ has a different shape when plotted. Indeed, the emitted power is minimal at periastron and maximal far away at the cosmological horizon, which is quite different from what one would expect in a scattering process. 
This illustrates a surprising and non-intuitive property of the de Sitter universe. 
Nonetheless, the contribution of $P_4H^4$ to the radiated energy is so small that it does not affect the total energy. 
As the $P_4H^4$ case, the shape of $P_3H^3$ presents asymptotes that take larger values close to the horizon bounds, which, however, are still negligible as compared to the leading de Sitter contributions, and, as aforementioned, the odd contributions do not contribute to the radiated energy as the integral vanishes.

Note that the static de Sitter universe has a horizon located at $r=h^{-1}$. The angles corresponding to the 
horizon are the bounds between which the above functions for the power are defined (thus  $\phi\in[\phi_-,\phi_+]$) and 
correspond to the bounds of the integral giving the total radiated energy:
\begin{equation}
    \frac{1}{h}=\frac{r_0(1+\epsilon)}{1+\epsilon\cos\phi_\pm}\quad\Leftrightarrow\quad\phi_\pm=\pm\arccos\bigg(\frac{hr_0(1+\epsilon)-1}{\epsilon}\bigg)~.
\end{equation}
The radiated energy is given by
\begin{equation}
    \Delta E=\int P\,dt=\int_{\phi_-}^{\phi_+}\frac{P}{\dot{\phi}}\,d\phi=\sum_{n=0}^4\,H^n\int_{\phi_-}^{\phi_+}\frac{P_n}{\dot{\phi}}\,d\phi:=\sum_{n=0}^4\,\Delta E_n\,H^n~.
\end{equation}
As already mentioned $P_1$ and $P_3$ are clearly odd with respect to $\phi=0$, hence the respective
integrals between $\phi_-$ and $\phi_+$ vanish. For the different energy contributions we get
\begin{equation} \label{exp_init}
    \begin{split}
        \Delta E_0&=\frac{G^\frac{7}{2}\mu^2m^\frac{5}{2}}{90c^5r_0^\frac{7}{2}(1+\epsilon)^\frac{7}{2}}\bigg[12\phi_+(96+292\epsilon^2+37\epsilon^4)+24\epsilon^2(71+12\epsilon^2)\sin(2\phi_+)+368\epsilon^3\sin(3\phi_+)\\
        &+\sqrt{(\epsilon+1)(1-hr_0)(\epsilon-1+hr_0+\epsilon hr_0)}\;(4608+3504\epsilon^2)+33\epsilon^4\sin(4\phi_+)\bigg]~, \\
        \Delta E_2&=\frac{16G^\frac{5}{2}m^\frac{3}{2}\mu^2}{15c^5\sqrt{r_0}}\bigg(\sqrt{(1-hr_0)(\epsilon-1+hr_0+\epsilon hr_0)}\;(19+\frac{9\epsilon-9}{hr_0}) \\
        &\hspace{3cm}+\frac{20\phi_+}{\sqrt{\epsilon+1}}+10\sqrt{\epsilon-1}\,\text{arctanh}\Big[\sqrt{\frac{\epsilon-1}{\epsilon+1}}\tan(\phi_+/2)\Big]\bigg)~, \\
        \Delta E_4&=\frac{16G^\frac{3}{2}\sqrt{m}\mu^2r_0^\frac{5}{2}}{15\,c^5}\bigg(\frac{\sqrt{\epsilon^2-(hr_0\epsilon+hr_0-1)^2}}{3h^3r_0^3(\epsilon-1)^2\sqrt{\epsilon+1}}\big[4-2hr_0-11h^2r_0^2+2\epsilon(hr_0-4) \\
        &\hspace{3cm}+\epsilon^2(4+5h^2r_0^2)\big]-2\frac{\epsilon^2-3}{(\epsilon-1)^\frac{5}{2}}\,\text{arctanh}\Big[\sqrt{\frac{\epsilon-1}{\epsilon+1}}\tan(\phi_+/2)\Big]\bigg)~,
    \end{split}
\end{equation}
where in the above formula we have inserted the speed of light $c$ dependence explicitly and used
$h=\frac{H}{c}$.

The expressions in (\ref{exp_init}) depend explicitly on the Hubble rate and also implicitly through $\phi_-$ and $\phi_+$.
As next we make an expansion in powers of $H$ (equivalently of $h$) of the above expressions such as to get all terms up to $H^2$. We neglect higher order terms as they are extremely small.

\paragraph{Expansion of $\Delta E_0$:}

For this leading term we use the following expansions: 
\begin{equation} \label{three_expansions}
    \begin{split}
        &\phi_+=\arccos\bigg(\frac{hr_0(1+\epsilon)-1}{\epsilon}\bigg)=\arccos\bigg(\frac{-1}{\epsilon}\bigg)-hr_0\sqrt{\frac{\epsilon+1}{\epsilon-1}}+\frac{h^2r_0^2}{2}\frac{\sqrt{\epsilon+1}}{(\epsilon-1)^\frac{3}{2}}+\mathcal{O}(h^3)~,\\
        &\sqrt{(1-hr_0)(\epsilon-1+hr_0+\epsilon hr_0)}=\sqrt{\epsilon-1}+\frac{hr_0}{\sqrt{\epsilon-1}}-\frac{h^2r_0^2}{2}\frac{\epsilon^2}{(\epsilon-1)^\frac{3}{2}}+\mathcal{O}(h^3)~, \\
        &\sin\phi_+=\frac{\sqrt{\epsilon^2-1}}{\epsilon}+\frac{hr_0}{\epsilon}\sqrt{\frac{\epsilon+1}{\epsilon-1}}-h^2r_0^2\frac{\epsilon^2}{2}\frac{\sqrt{\epsilon+1}}{(\epsilon-1)^\frac{3}{2}}+\mathcal{O}(h^3)~,
    \end{split}
\end{equation}
and similarly for $\sin(n\phi_+)$, $n=2,3,4$, whose expressions for the expansions are given in appendix \ref{sinus_expansions_derivation}. By plugging these expansions into $\Delta E_0$ of (\ref{exp_init}) and gathering together the terms proportional to the same power of $H$ (or $h$), we get the final expanded expression for $\Delta E_0$:
\begin{equation} \label{final_DE0_M2}
    \Delta E_0=\frac{2G^\frac{7}{2}\mu^2m^\frac{5}{2}}{45c^5r_0^\frac{7}{2}(\epsilon+1)^\frac{7}{2}}\bigg(3\arccos\Big(\frac{-1}{\epsilon}\Big)(96+292\epsilon^2+37\epsilon^4)+\sqrt{\epsilon^2-1}\,(602+673\epsilon^2)\bigg)+\mathcal{O}(H^3)~.
\end{equation}
Note that there are no terms proportional to $H$ and $H^2$. As expected, this leading term corresponds to the Minkowski contribution already computed in (\ref{total_energy_Mink_method2}).

\paragraph{Expansion of $\Delta E_2H^2$:}
This term corresponds to the leading de Sitter contribution. The expansion of $\tan(\phi_+/2)$ is given in appendix \ref{sinus_expansions_derivation}, equation (\ref{tan_horrible}) so that we get:
\begin{equation} 
    \sqrt{\frac{\epsilon-1}{\epsilon+1}}\tan(\frac{\phi_+}{2})=1-\frac{\epsilon}{\epsilon-1}hr_0+\frac{\epsilon^2}{(\epsilon-1)^2}\frac{h^2r_0^2}{2}+\mathcal{O}(h^3)~,
\end{equation}
and as next we have to expand the arctanh of this expression, for which we take the expansion for an argument around 1
as follows:
\begin{equation}
    \text{arctanh}(1-x)=-\frac{1}{2}\ln\big(\frac{x}{2}\big)-\frac{x}{4}-\frac{x^2}{16}+\mathcal{O}(x^3)\quad\text{for}\quad 0<x\ll1~.
\end{equation}
With the two above equations we get the desired expansion of the arctanh term:
\begin{equation}
    \text{arctanh}\Big[\sqrt{\frac{\epsilon-1}{\epsilon+1}}\tan(\frac{\phi_+}{2})\Big]=-\frac{1}{2}\ln\Big(\frac{\epsilon}{\epsilon-1}\frac{hr_0}{2}+\mathcal{O}(h^2)\Big)-\frac{\epsilon}{\epsilon-1}\frac{hr_0}{4}+\frac{\epsilon^2}{(\epsilon-1)^2}\frac{h^2r_0^2}{16}+\mathcal{O}(h^3)~.
\end{equation}
Finally, the quadratic contribution term can be simplified and rewritten as
\begin{equation}\label{final_DE2_M2}
    \begin{split}
        \Delta E_2H^2&=\frac{48G^\frac{5}{2}\mu^2m^\frac{3}{2}(\epsilon-1)^\frac{3}{2}}{5c^4r_0^\frac{3}{2}}\,H \\
        &+\frac{16G^\frac{5}{2}\mu^2m^\frac{3}{2}\,H^2}{15c^5\sqrt{r_0}}\bigg(\frac{20\arccos(-1/\epsilon)}{\sqrt{\epsilon+1}}-5\sqrt{\epsilon-1}\,\ln\Big(\frac{\epsilon}{\epsilon-1}\frac{Hr_0}{2c}+\mathcal{O}(H^2)\Big)+28\sqrt{\epsilon-1}\bigg)~.
    \end{split}
\end{equation}

\paragraph{Expansion of $\Delta E_4H^4$:}
Although at first glance it is proportional to $H^4$, due to the $H$ dependence in the limiting angles, we get
terms proportional to $H$ and $H^2$. Clearly these are very small terms.
Indeed the expansion of the square root term gives:
\begin{equation} \label{second_sqrt_expansion}
    \sqrt{\epsilon^2-(1-hr_0(\epsilon+1))^2}=\sqrt{\epsilon^2-1}+hr_0\sqrt{\frac{\epsilon+1}{\epsilon-1}}-\frac{h^2r_0^2}{2}\frac{\epsilon^2\sqrt{\epsilon+1}}{(\epsilon-1)^\frac{3}{2}}+\mathcal{O}(h^3)~.
\end{equation}
By plugging expansion (\ref{second_sqrt_expansion}) into $\Delta E_4$ in (\ref{exp_init}), regrouping the terms according to their respective orders and keeping only terms up to quadratic order in $H$, we get:
\begin{equation}\label{final_DE4_M2}
    \begin{split}
        \Delta E_4H^4&=\frac{64G^\frac{3}{2}\mu^2\sqrt{m}}{45c^2\sqrt{r_0}}\sqrt{\epsilon-1}\,H+\frac{32G^\frac{3}{2}\mu^2\sqrt{mr_0}}{15c^3\sqrt{\epsilon-1}}\,H^2+\mathcal{O}(H^3)~.
    \end{split}
\end{equation}

\paragraph{Final expansion:}
We can now write the total radiated energy up to $H^2$ terms by adding the contributions as given in (\ref{final_DE0_M2}), (\ref{final_DE2_M2}) and (\ref{final_DE4_M2}). This way we get
\begin{equation} \label{coloc_Momo}
    {\Delta E=\epsilon_0+\epsilon_1H-H^2\frac{16G^\frac{5}{2}m^\frac{3}{2}\mu^2\sqrt{\epsilon-1}}{3c^5\sqrt{r_0}}\ln\Big[\frac{\epsilon}{\epsilon-1}\frac{Hr_0}{2c}+\mathcal{O}(H^2)\Big]+\epsilon_2H^2+\mathcal{O}(H^3)}
\end{equation}
with
\begin{equation}
    \epsilon_0=\frac{2G^\frac{7}{2}\mu^2m^\frac{5}{2}}{45c^5r_0^\frac{7}{2}(\epsilon+1)^\frac{7}{2}}\bigg(3\arccos\Big(\frac{-1}{\epsilon}\Big)(96+292\epsilon^2+37\epsilon^4)+\sqrt{\epsilon^2-1}\,(602+673\epsilon^2)\bigg)~,
\end{equation}
which is the leading term corresponding to the Minkowski contribution already computed in (\ref{total_energy_Mink_method2}). The leading de Sitter contribution is given by:
\begin{equation}
    \epsilon_1=\frac{48G^\frac{5}{2}\mu^2m^\frac{3}{2}}{5c^4r_0^\frac{3}{2}}(\epsilon-1)^\frac{3}{2}+\frac{64G^\frac{3}{2}\mu^2\sqrt{m}}{45c^2\sqrt{r_0}}\sqrt{\epsilon-1}~.
\end{equation}
Finally, the leading quadratic $H^2$ contribution reads:
\begin{equation}
    \begin{split}
        \epsilon_2&=\:\frac{32G^\frac{3}{2}\mu^2\sqrt{mr_0}}{15c^3\sqrt{\epsilon-1}}+\frac{64G^\frac{5}{2}m^\frac{3}{2}\mu^2}{15c^5\sqrt{r_0}}\bigg(7\sqrt{\epsilon-1}+\frac{5}{\sqrt{\epsilon+1}}\arccos\Big(\frac{-1}{\epsilon}\Big)\bigg)~.
    \end{split}
\end{equation}
(Terms proportional to $H^2$ ln$\,H$ are also neglected.)

\paragraph{Numerical application:} To get a better idea of the magnitude of the various terms discussed above we consider as an example (given also in Fig. \ref{power_plot_NA1}) an hyperbolic encounter of two supermassive black holes, which have the same mass $m_1=m_2=0.5\cdot10^{7} M_\odot$, with an impact parameter $b=10 AU$ and an initial relative velocity $v_0=10^7 m/s$. 
The de Sitter universe is characterized by a cosmological constant whose value is $\Lambda=1.1056\cdot10^{-52}m^{-2}$. 
In table \ref{M1_NA1} we give the numerical values we get, using our example, for the above contributions to the total emitted energy
(expressed in units of Joule).

\begin{table}[h!]
    \centering
    \begin{tabular}{|c|c|c|c|}
    \hline
    $\epsilon_0$ & $\epsilon_1\,H$ & $\epsilon_2\,H^2$ & const$\cdot H^2\ln{H}$\\
    \hline
    \hline
    $3.9\cdot10^{51}$ & $4.8\cdot10^{35}$ & $7.5\cdot10^{22}$ & $4.9\cdot10^{21}$ \\
    \hline
    \end{tabular}
    \caption{Total radiated energy in Joule during a hyperbolic encounter of two supermassive black holes whose single scattering event is characterised by their masses $m_1=m_2=0.5\cdot10^{7} M_\odot$, their impact parameter $b=10 AU$ and their initial relative velocity $v_0=10^7 m/s$; both in a Minkowskian and de Sitter universe. The different contributions are displayed. 
    (Notice that $b=r_0\sqrt{\frac{\epsilon +1}{\epsilon-1}}$ and $v_0^2 = \frac{Gm}{r_0}(\epsilon-1)$)}
    \label{M1_NA1}
\end{table}

From this example we see that the leading de Sitter term (proportional to $H$) is extremely small as compared to the main Minkowski contribution. Higher terms are even smaller, and thus negligible. 
This conclusion also holds assuming other values for the masses, impact parameters, and velocities.
Clearly, with present or planned detectors for gravitational waves such an effect is too small to be observable (see also Section 5 of \cite{naf} for a discussion on this issue).

\paragraph{Parabolic limit:}
We discuss now the parabolic limit ($\epsilon \rightarrow 1$). A detailed analysis shows that the exact parabolic limit is only reachable when $H \rightarrow 0$, recovering thus the Minkowski
spacetime. Note that we have $\epsilon>1$ and $\epsilon \rightarrow 1$
\begin{equation}
    h=\frac{1+\epsilon\cos\phi_\pm}{r_0(1+\epsilon)}\approx\frac{1-\epsilon}{r_0(1+\epsilon)}\approx\frac{1-\epsilon}{2r_0}~.
\end{equation}
However, as we are considering a positive Hubble rate together with an eccentricity close to but higher than one, we thus 
get for the Hubble rate 
\begin{equation} \label{Hubble_rate_para_lim}
    H=hc=\frac{c}{2r_0}(\epsilon-1)
\end{equation}
for $\epsilon \rightarrow 1$ in the parabolic limit. As
expected the Hubble rate tends toward 0 as the orbit tends toward a parabola.
It is now straightforward to perform the limit in the above expressions for the energy, so that we find
that in the parabolic limit we get only the Minkowski expression 
\begin{equation}
    \lim_{\epsilon\to1}\Delta E=\frac{85\pi\,G^\frac{7}{2}\mu^2m^\frac{5}{2}}{12\sqrt{2}\,c^5r_0^{7/2}}~,
\end{equation}
which we found in eq. (\ref{parabolic}).

\section{Conclusion} \label{conclusion}
We used the de Sitter quadrupole formula expressing the power radiated in the form of gravitational waves by a gravitationally interacting system in an exponentially expanding universe for the particular case of a binary hyperbolic encounter. For the radiated energy we used an expansion in powers of the Hubble rate $H$. It turns out that the leading de Sitter contribution proportional to $H$ is extremely small and thus not measurable with the present and future planned gravitational wave observatories and will not cause biases in the parameter estimation conducted with waveform assuming an asymptotically flat background. We also checked that the parabolic limit is consistent with the value found in Minkowski space since indeed this limit exists only within the framework of it. Here we did not take into account the shrinking and expansion of the trajectory due to the loss of gravitational radiation and the expansion of the universe during the encounter, respectively. It would be of interest to study the evolution of orbital parameters through these two factors. 

In this work, we have not discussed the implication that the de Sitter background has on the linear memory of the hyperbolic encounters; this can be found in \cite{compere}. It would be interesting to relate this study to a deeper understanding of the memory effect in de Sitter background for hyperbolic encounters. 

\section*{Acknowledgements} 
PJ is supported by the Swiss Space Office, Bern, and ST is supported by the Swiss National Science Foundation Ambizione Grant Number : PZ00P2-202204.  

\appendix

\section{Derivatives of moments}
The computations of the radiated power need time derivatives of the components of the quadrupole moment tensor. In the Minkowski case, only the third time derivative is needed, whereas the de Sitter case involves the first, second, and third time derivatives. In
this appendix, we give all the expressions of the time derivatives of the moments.

\subsection{Time derivatives } \label{DOM_M2}

\begin{equation}
    \begin{split}
        \dot{Q}_{11}&=-\mu L\,\frac{\sin(2\phi)}{1+\epsilon\cos\phi} \\
         \dot{Q}_{12}&=\mu L\,\frac{\epsilon\cos\phi+\cos(2\phi)}{1+\epsilon\cos\phi} \\
         \dot{Q}_{22}&=2\mu L\,\frac{\epsilon+\cos\phi}{1+\epsilon\cos\phi}\sin\phi 
    \end{split}
\end{equation}

\begin{equation}
    \begin{split}
        \ddot{Q}_{11}&=\frac{-G\mu m}{2r_0(\epsilon+1)}\Big(3\epsilon\cos\phi+4\cos(2\phi)+\epsilon\cos(3\phi)\Big) \\
         \ddot{Q}_{12}&=\frac{-G\mu m}{r_0(\epsilon+1)}\Big(4\cos\phi+3\epsilon+\epsilon\cos(2\phi)\Big)\sin\phi \\
         \ddot{Q}_{22}&=\frac{G\mu m}{2r_0(\epsilon+1)}\Big(7\epsilon\cos\phi+4\cos(2\phi)+4\epsilon^2+\epsilon\cos(3\phi)\Big)
    \end{split}
\end{equation}

\begin{equation}
    \begin{split}
        \dddot{Q}_{11}&=\frac{\mu L^3}{r_0^4(\epsilon+1)^4}(1+\epsilon\cos\phi)^2(4+3\epsilon\cos\phi)\sin(2\phi) \\
        \dddot{Q}_{12}&=\frac{-\mu L^3}{2r_0^4(\epsilon+1)^4}(1+\epsilon\cos\phi)^2\Big(5\epsilon\cos\phi+8\cos(2\phi)+3\epsilon\cos(3\phi)\Big) \\
        \dddot{Q}_{22}&=\frac{-\mu L^3}{r_0^4(\epsilon+1)^4}(1+\epsilon\cos\phi)^2\Big(8\cos\phi+5\epsilon+3\epsilon\cos(2\phi)\Big)\sin\phi
    \end{split}
\end{equation}
where $L=\sqrt{Gmr_0(\epsilon+1)}$ stands for the angular momentum per unit mass of the system, which is a time conserved quantity.

\section{Trigonometric functions expansions} \label{sinus_expansions_derivation}

In this appendix, we provide the trigonometric relations needed for the expansion in powers of $H=hc$ of the formulas for the total radiated energy. We used
\begin{equation}
    \phi_+=\arccos\bigg(\frac{hr_0(1+\epsilon)-1}{\epsilon}\bigg)=\arccos\bigg(\frac{-1}{\epsilon}\bigg)-hr_0\sqrt{\frac{\epsilon+1}{\epsilon-1}}+\frac{h^2r_0^2}{2}\frac{\sqrt{\epsilon+1}}{(\epsilon-1)^\frac{3}{2}}+\mathcal{O}(h^3)~.
\end{equation}
With the following expansion,
\begin{equation}
    \sin(a+x)=\sin a+\cos a\,x-\sin a\,\frac{x^2}{2}+\mathcal{O}(x^3)
\end{equation}
we infer the expansion of $\sin(n\phi_+)$ for $n\geq0$:
\begin{equation} \label{waterloo_ABBA}
    \begin{split}
        \sin(n\phi_+)&=\sin\Big(n\arccos\big(\frac{-1}{\epsilon}\big)\Big)-hr_0n\sqrt{\frac{\epsilon+1}{\epsilon-1}}\cos\Big(n\arccos\big(\frac{-1}{\epsilon}\big)\Big) \\
        &+h^2r_0^2\frac{n}{2}\frac{\sqrt{\epsilon+1}}{(\epsilon-1)^\frac{3}{2}}\bigg[\cos\Big(n\arccos\big(\frac{-1}{\epsilon}\big)\Big)-n\sqrt{\epsilon^2-1}\sin\Big(n\arccos\big(\frac{-1}{\epsilon}\big)\Big)\bigg] \\
        &+\mathcal{O}(h^3)~.
    \end{split}
\end{equation}
This expansions has terms of the form $\sin\Big(n\arccos\big(\frac{-1}{\epsilon}\big)\Big)$ and $\cos\Big(n\arccos\big(\frac{-1}{\epsilon}\big)\Big)$ which can be simplified using trigonometric relations:
\begin{equation} \label{touslescosetsinphi0}
    \begin{split}
        \sin\Big(\arccos\big(\frac{-1}{\epsilon}\big)\Big)&=\frac{\sqrt{\epsilon^2-1}}{\epsilon} \\
        \sin\Big(2\arccos\big(\frac{-1}{\epsilon}\big)\Big)&=-\frac{2}{\epsilon^2}\sqrt{\epsilon^2-1} \\
        \sin\Big(3\arccos\big(\frac{-1}{\epsilon}\big)\Big)&=\sqrt{\epsilon^2-1}\big(\frac{4}{\epsilon^3}-\frac{1}{\epsilon}\big) \\
        \sin\Big(4\arccos\big(\frac{-1}{\epsilon}\big)\Big)&=4\sqrt{\epsilon^2-1}\big(\frac{1}{\epsilon^2}-\frac{2}{\epsilon^4}\big) \\
        \cos\Big(\arccos\big(\frac{-1}{\epsilon}\big)\Big)&=\frac{-1}{\epsilon} \\
        \cos\Big(2\arccos\big(\frac{-1}{\epsilon}\big)\Big)&=\frac{2}{\epsilon^2}-1 \\
        \cos\Big(3\arccos\big(\frac{-1}{\epsilon}\big)\Big)&=\frac{3}{\epsilon}-\frac{4}{\epsilon^3} \\
        \cos\Big(4\arccos\big(\frac{-1}{\epsilon}\big)\Big)&=\frac{8}{\epsilon^4}-\frac{8}{\epsilon^2}+1~,
    \end{split}
\end{equation}
which can be plugged into (\ref{waterloo_ABBA}) in order to obtain the desired results:
\begin{equation}
    \begin{split}
        \sin\phi_+&=\frac{\sqrt{\epsilon^2-1}}{\epsilon}+\frac{hr_0}{\epsilon}\sqrt{\frac{\epsilon+1}{\epsilon-1}}-h^2r_0^2\frac{\epsilon^2}{2}\frac{\sqrt{\epsilon+1}}{(\epsilon-1)^\frac{3}{2}}+\mathcal{O}(h^3) \\
        \sin(2\phi_+)&=-\frac{2}{\epsilon^2}\sqrt{\epsilon^2-1}-2hr_0\sqrt{\frac{\epsilon+1}{\epsilon-1}}\big(\frac{2}{\epsilon^2}-1\big)+h^2r_0^2\,\frac{3\epsilon^2-2}{\epsilon^2(\epsilon-1)^{\frac{3}{2}}}\sqrt{\epsilon+1}+\mathcal{O}(h^3) \\
        \sin(3\phi_+)&=\big(\frac{4}{\epsilon^3}-\frac{1}{\epsilon}\big)\sqrt{\epsilon^2-1}-3hr_0\sqrt{\frac{\epsilon+1}{\epsilon-1}}\big(\frac{3}{\epsilon}-\frac{4}{\epsilon^3}\big)+\frac{3}{2}h^2r_0^2\frac{\sqrt{\epsilon+1}}{\epsilon^3(\epsilon-1)^\frac{3}{2}}(3\epsilon^4-12\epsilon^2+8)+\mathcal{O}(h^3) \\
        \sin(4\phi_+)&=4\big(\frac{1}{\epsilon^2}-\frac{2}{\epsilon^4}\big)\sqrt{\epsilon^2-1}-4hr_0\sqrt{\frac{\epsilon+1}{\epsilon-1}}\big(1+\frac{8}{\epsilon^4}-\frac{8}{\epsilon^2}\big) \\
        &+h^2r_0^2\frac{2}{\epsilon^4}\frac{\sqrt{\epsilon+1}}{(\epsilon-1)^\frac{3}{2}}(-15\epsilon^4+40\epsilon^2-24)+\mathcal{O}(h^3)~.
    \end{split}
\end{equation}

We needed also the expansion of $\tan(\phi_+/2)$. We again resort to the general expansion of the tangent function:
\begin{equation}
    \tan(a+x)=\tan a+\frac{x}{\cos^2a}+\frac{\tan a}{\cos^2a}x^2+\mathcal{O}(x^3)~.
\end{equation}
After some algebra, this leads us to:
\begin{equation}
    \tan\frac{\phi_+}{2}=\tan\frac{\phi_0}{2}-\frac{hr_0}{2\cos^2(\phi_0/2)}\sqrt{\frac{\epsilon+1}{\epsilon-1}}+\frac{h^2r_0^2}{4\cos^2(\phi_0/2)}\frac{\sqrt{\epsilon+1}}{(\epsilon-1)^\frac{3}{2}}\Big(1+\tan(\frac{\phi_0}{2})\sqrt{\epsilon^2-1}\Big)+\mathcal{O}(h^3)~.
\end{equation}
Furthermore, recalling that $\phi_0$ is linked to $\epsilon$ via $\cos(\phi_0)=-1/\epsilon$, we arrive at the useful equation:
\begin{equation} \label{tan_horrible}
    \sqrt{\frac{\epsilon-1}{\epsilon+1}}\tan(\frac{\phi_+}{2})=1-\frac{\epsilon}{\epsilon-1}hr_0+\frac{\epsilon^2}{(\epsilon-1)^2}\frac{h^2r_0^2}{2}+\mathcal{O}(h^3)~.
\end{equation}

\section*{References}
\bibliographystyle{unsrt}
\bibliography{mybib}

\begin{thebibliography}{10}

\bibitem{gwtc-3}
R.~Abbott et~al.
\newblock {GWTC-3: Compact Binary Coalescences Observed by LIGO and Virgo during the Second Part of the Third Observing Run}.
\newblock {\em Phys. Rev. X}, 13(4):041039, 2023.

\bibitem{ligo}
J.~Aasi et~al.
\newblock {Advanced LIGO}.
\newblock {\em Class. Quant. Grav.}, 32:074001, 2015.

\bibitem{virgo}
F.~Acernese et~al.
\newblock {Advanced Virgo: a second-generation interferometric gravitational wave detector}.
\newblock {\em Class. Quant. Grav.}, 32(2):024001, 2015.

\bibitem{kagra}
Yoichi Aso, Yuta Michimura, Kentaro Somiya, Masaki Ando, Osamu Miyakawa, Takanori Sekiguchi, Daisuke Tatsumi, and Hiroaki Yamamoto.
\newblock Interferometer design of the kagra gravitational wave detector.
\newblock {\em Phys. Rev. D}, 88:043007, Aug 2013.

\bibitem{Sahni:1999gb}
Varun Sahni and Alexei~A. Starobinsky.
\newblock {The Case for a positive cosmological Lambda term}.
\newblock {\em Int. J. Mod. Phys. D}, 9:373--444, 2000.

\bibitem{zp-hype}
Ya.~B. {Zel'dovich} and A.~G. {Polnarev}.
\newblock {Radiation of gravitational waves by a cluster of superdense stars}.
\newblock {\em J Soviet Astronomy}, 18:17, August 1974.

\bibitem{gb-hype}
Juan Garc\'\i{}a-Bellido and Savvas Nesseris.
\newblock {Gravitational wave energy emission and detection rates of Primordial Black Hole hyperbolic encounters}.
\newblock {\em Phys. Dark Univ.}, 21:61--69, 2018.

\bibitem{cap-hype}
SALVATORE CAPOZZIELLO, MARIAFELICIA DE~LAURENTIS, FRANCESCO DE~PAOLIS, G.~INGROSSO, and ACHILLE NUCITA.
\newblock Gravitational waves from hyperbolic encounters.
\newblock {\em Modern Physics Letters A}, 23(02):99--107, 2008.

\bibitem{bong-hype}
B\'eatrice Bonga and Jeffrey~S. Hazboun.
\newblock {Power radiated by a binary system in a de Sitter Universe}.
\newblock {\em Phys. Rev. D}, 96(6):064018, 2017.

\bibitem{ho-hype}
Sk~Jahanur Hoque and Ankit Aggarwal.
\newblock Quadrupolar power radiation by a binary system in de sitter background.
\newblock {\em International Journal of Modern Physics D}, 28(01):1950025, 2019.

\bibitem{Turner_1977}
M.~{Turner}.
\newblock {Gravitational radiation from point-masses in unbound orbits: Newtonian results.}
\newblock {\em The Astrophysical Journal}, 216:610--619, September 1977.

\bibitem{Hoque:2018byx}
Sk~Jahanur Hoque and Amitabh Virmani.
\newblock {On Propagation of Energy Flux in de Sitter Spacetime}.
\newblock {\em Gen. Rel. Grav.}, 50(4):40, 2018.

\bibitem{Date:2015kma}
Ghanashyam Date and Sk~Jahanur Hoque.
\newblock {Gravitational waves from compact sources in a de Sitter background}.
\newblock {\em Phys. Rev. D}, 94(6):064039, 2016.

\bibitem{naf}
Joachim Naf, Philippe Jetzer, and Mauro Sereno.
\newblock {On Gravitational Waves in Spacetimes with a Nonvanishing Cosmological Constant}.
\newblock {\em Phys. Rev. D}, 79:024014, 2009.

\bibitem{compere}
Geoffrey Comp\`ere, Sk~Jahanur Hoque, and Emine~\c{S}eyma Kutluk.
\newblock {Quadrupolar radiation in de Sitter: displacement memory and Bondi metric}.
\newblock {\em Class. Quant. Grav.}, 41(15):155006, 2024.

\end{thebibliography}

\end{document}